# Kinetic Monte Carlo Simulation of Single-Electron Multiple-Trapping Transport in Disordered Media

Mohammad Javadi and Yaser Abdi*

*Department of Physics, University of Tehran, Tehran 14395-547, Iran*

Abstract

The conventional single-particle Monte Carlo simulation of charge transport in disordered media is based on the truncated density of localized states (DOLS) which benefits from very short time execution. Although this model successfully clarifies the properties of electron transport in moderately disordered media, it overestimates the electron diffusion coefficient for strongly disordered media. The origin of this deviation is discussed in terms of zero-temperature approximation in the truncated DOLS and the ignorance of spatial occupation of localized states. Here, based on the multiple-trapping regime we introduce a modified single-particle kinetic Monte Carlo model that can be used to investigate the electron transport in any disordered media independent from the value of disorder parameter. In the proposed model, instead of using a truncated DOLS we imply the raw DOLS. In addition, we have introduced an occupation index for localized states to consider the effect of spatial occupation of trap sites. The proposed model is justified in a simple cubic lattice of trap sites for broad interval of disorder parameters, Fermi levels, and temperatures.

---

*Corresponding author. Tel/Fax: +98 21 61118610, E-mail address: y.abdi@ut.ac.ir

# I. INTRODUCTION

Inorganic nano-crystalline metal-oxide materials such as $TiO_2$ and ZnO have been drawing a lot of attention due to photovoltaic and photocatalytic applications.[1, 2] Considering the key role of charge transport in these applications [3], description and modeling of charge transport phenomenon within such materials is very important. Because of their nano-sized crystalline structure, the charge transport in this class of metal-oxides essentially differs from crystalline counterparts. Indeed, in the presence of structural disorder some extended electronic states become localized. At intermediate disorder, which is approximately fulfilled in the nano-crystalline inorganic materials, localized and extended states coexist at different energies. The states in the middle of the band remain extended and separated from localized states near the band tail by means of mobility edge. In the case of inorganic noncrystalline materials, the density of localized states (DOLS) is given by [4]

$$g(E) = \alpha \frac{N_t}{kT} exp\left(-\alpha \frac{E}{kT}\right) \qquad (1)$$

in this equation $k$ and $T$ are respectively, the Boltzmann constant and the absolute temperature. $N_t$ is total density of localized states and $E$ is the energy of localized states measured with respect to the mobility edge as reference level ($E_l$). The parameter $\alpha$ indicates the measure of disorder in the energy distribution of localized states. This parameter is related to the characteristic energy ($E_0$) and temperature ($T_0$) of localized states via $\alpha = kT/E_0 = T/T_0$.[4] Generally, the value of $\alpha$ resides between zero and one. A smaller value of alpha means a broader distribution of the energies of localized states (i.e. stronger disorder in the energy distribution).

Multiple-trapping (MT) model is often applied to investigate the charge transport in the noncrystalline inorganic semiconductors due to its successful interpretation of dispersive or non-Gaussian transport.[4-6] This model is used when the localized states and extended states coexist in the system. The model assumes that the electron transport is carried out in the extended states. However, when the electron encounters a localized state (which acts as a trap), it falls down into that trap and stays there until it thermally activated to the extended state where it is again free to continue its motion. In other words, in the framework of MT model it is assumed that the electrons are immobile in the localized states and the charge transport is solely carried out via extended states (see Figure 1). In this regard, the motion of electrons in the extended states is slowed down by successive trapping/detrapping events induced by localized states (traps). In addition, in the MT model the direct transition between localized states is ignored since at high enough temperature the transition rate to the extended states is much higher than the tunneling rate between localized states [4, 7]. The MT model differs from the Hopping model, where the transport mechanism is based on the direct transition or tunneling between localized states (the transition rate is exponentially related to the energy difference between initial and final states as well as the distance between them).[4] The hopping model is applied to the organic semiconductors where all states are localized (there are no extended states or bands). It is also applied to the inorganic semiconductors (where the localized and delocalized states coexist) when the temperature is low enough. In this work, we focus on the Monte Carlo simulation of MT transport in the inorganic semiconductors.

Computer simulation of charge transport in disordered media is based on the continuous-time random walk which offers a comprehensive approach for investigation of different morphological and energetic aspects of electrons transport.[8-11] The kinetic Monte Carlo can be used to simulate the time evolution of the processes occurring with known transition rates.[12] In order to investigate the electron transport at different Fermi levels, two alternative approaches based on the many-particle Monte Carlo (MC) simulation have been developed. In the first approach the Fermi level is considered as an input parameter by tuning the ratio of electrons to the number of trap sites.[9, 11] In the second approach, the Fermi level is counted as a well-defined output parameter which is calculated from the ratio of visited states to the $g(E)$.[13, 14] The common disadvantage of both models is the amount of time pending for the simulation execution leading to an increasing trend toward the single-particle random walk.[15-21]

MC simulations show that in the absence of cross-correlation, the kinetic diffusion coefficient (which reflects the diffusion of the center of mass of N particles) and tracer diffusion coefficient (which reflects the diffusion of single particle) are essentially identical in a broad interval of carrier densities and temperatures.[22] Moreover, the electron-electron interactions can be ignored when the density of charge carriers is less than the total density of localized states. In this regard, the single-particle MC simulation with the benefit of short time implementation can be used.

The conventional single-particle random walk simulation of charge transport is based on a truncated DOLS as [14]

$$g(E) = \mathrm{H}(E - E_f)\alpha \frac{N_t}{kT} exp\left(-\alpha \frac{E}{kT}\right) \qquad (2)$$

where $H$ is the Heaviside step function and $E_f$ is the Fermi energy(see Figure 2). According to this equation, in the single-particle random walk simulation only localized states with energy higher than the Fermi level are considered for trapping process. This modification is relied upon the fact that for a reasonable period of time, the charge transport properties are governed by Fermi electrons.[11] It is shown that the single-particle simulation can approximately reproduce the many-particle results leading to a growing tendency toward the single-particle random walk.[14-21] In this regard, it is necessary to discuss a number of important matters in the conventional single-particle charge transport simulation in disordered media.

First, it seems that introducing the Heaviside step function in Eq.2 is based on the zero-temperature approximation for occupation probability of localized states where the Fermi-Dirac distribution ($f_{FD}(E.T)$) is replaced by a step function (i.e. $f_{FD}(E.T = 0)$ see Figure 2). However, in the framework of MT model, the transport mechanism is often assumed to be thermally activated (i.e. the conductance vanishes as $T \rightarrow 0$). In this condition, applying Eq.2 for thermally-activated MT appears to not be self-consistent. Moreover, as it will be shown in the subsequent, using the truncated DOLS we cannot predict the correct transport activation energy.

Second concern is about the spatial occupation of localized states. In the conventional single-particle random walk we always deal with a fixed lattice points of unoccupied localized states whose energies are distributed according to Eq.2. This means that the average distance between unoccupied sites is assumed to be constant and independent from temperature, Fermi level, and disorder parameter (see Figure 3). Assuming a homogeneous distribution of localized states, the average distance between unoccupied sites (localized states) can be obtained as $a_l = N_{imp}^{-1/3}$ where $N_{imp}$ is the volume density of unoccupied states and can be calculated as

$$N_{imp} = \int_{-\infty}^{E_l=0} g(E)\big(1 - f_{FD}(E.T)\big)dE \qquad (3)$$

substituting $g(E)$ from Eq. 1 and $f_{FD} = 1/(1 + exp(E - E_f\ )/kT)$ into the above equation we obtain

$$N_{imp} = \frac{\alpha}{1+\alpha} N_t\, e^{-\varepsilon_f}\ {}_2F_1(1.1 + \alpha.2 + \alpha. -e^{-\varepsilon_f}) \qquad (4)$$

here ${}_2F_1$ is the Gauss hypergeometric function and $\varepsilon_f$ is the normalized Fermi energy ($\varepsilon_f = E_f/kT$). Figure 4 shows the fraction of unoccupied localized states and the average distance between them as a function of normalized Fermi level. It is seen that in contrast of conventional single-particle model, $a_l$ and $N_{imp}$ are very sensitive to the position of Fermi level. In addition, as $\alpha$ decreases the average distance between unoccupied sites raises which signifies the region where the conventional single-particle model faces difficulties.

Third, the electron transport time in the extended states is often neglected in the simulation process since it is too small when compared with the electron residence time in the localized states.[14, 16, 18, 19, 21] However, we will show that considering the transport time in the extended states is essential when we add the spatial occupation of localized states to the simulation model.

Due to these issues the conventional single-particle MC overestimates the electron diffusion coefficient especially at lower values of $\alpha$ (strongly disordered media). Here, our aim is to propose a modified single-particle random walk for charge transport in disordered media. The proposed model eliminates the above mentioned concerns about the conventional single-particle model. In order to justify the modified model, we compare the results of MC simulations with the numerical calculations for electron diffusion coefficient.

# II. SIMULATION AND THEORY

### Conventional single-particle MC simulation (model #1)

A typical MC simulation of electron transport is performed in three steps: spatial distribution of localized states, energy distribution of these states, and the random walk process (see Figure 3). If the study does not consider the geometric aspect of traps distribution, then the spatial distribution of traps often performed in a simple cubic lattice where each grid (site) in the lattice indicates a localized state or trap. At a desired Fermi level, the energy of each localized state is calculated according to the Eq. 2 as $E_i = -(kT_0)Log(R)$ with the condition that $E_i > E_f$ ($R$ is a random number uniformly distributed between zero and one). For thermally-activated transport the electron residence time in a trap with energy $E_i$ is given by [4]

$$t_i = -ln(R)\nu_0^{-1}exp(-\frac{E_i - E_l}{kT}) \qquad (5)$$

here $\nu_0$ is the attempt-to-scape frequency, assumed equal to the lattice vibrations frequency (1-10 THz). The random pre-factor is used for considering stochastic fluctuations. The random walk process begins by placing an electron randomly at one of the trap sites (step III in Figure 3). Then the simulation time is advanced by residence time of that site and the electron is allowed to move randomly into one the nearest neighbors. The electron transport time in the extended states is often ignored in the simulation process because it is very small when compared with $t_i$ ($\tau \ll \langle t_i \rangle$). [14, 16, 18, 19, 21] In the thermal equilibrium, the electron transport exhibits non-dispersive character (i.e. after sufficiently long times) [6, 23] and the jump (or tracer) diffusion coefficient can be calculated as

$$D_j = \frac{\langle r^2(t) \rangle}{6t} \qquad (6)$$

where $\langle r^2(t) \rangle$ is mean-squared displacement and t is the total transport time. In order to investigate the effect of Fermi level on the charge transport, the energies of traps are redistributed with respect to Eq.2 and the new $E_f$ at the same lattice of trap sites and the simulation is repeated. It is noted that in this model, we always deal with an electron wander in an entirely unoccupied lattice of trap sites.

**Modified single-particle MC simulation (model #2)**

In this model the energies of trap sites are computed according to the original DOLS Eq. 1 (i.e. $E_i = -(kT_0)Log(R)$ ). Unlike model #1, the traps energies are not redistributed for new Fermi level. Instead we introduce an occupation index for each trap. The wanderer electron cannot occupy a blocked trap whose occupation index is one. When the electron come across the trap $i$ the transport time is calculated according to

$$t_i = -(1 - F_i)ln(R)\nu_0^{-1}exp(\frac{E_i - E_l}{kT}) + \tau \qquad (7)$$

The first term of this equation is the electron residence time (Eq. 5) with an additional term $(1 - F_i)$ which is introduced for considering the effect of spatial occupation of localized states. The index $F_i$ indicates the probability of occupancy of the trap $i$. When the electron come across an occupied site ($F_i = 1$) it will not fall into the trap and we do not need to consider a time for detrapping. Otherwise, i.e. when the electron encounters with an unoccupied site ($F_i = 0$), the electron will be trapped in that site and the residence time should be added to the transport time. In this regard, the trapping/detrapping events are taken into account only for the unoccupied localized states (Figure 2.c and Figure 3.c). At the desired Fermi level, the occupation indices of trap sites are calculated according to the Fermi-Dirac occupation statistics

$$F_i = \begin{cases} 0 & R > f_{FD}(E_i - E_f) \\ 1 & R \leq f_{FD}(E_i - E_f) \end{cases} \qquad (8)$$

again $R$ is a random number uniformly distributed between zero and one. It is noted that by this modification the spatial occupation of localized states is considered in the simulation.

The second term in Eq. 7 ($\tau$) is the average time of transport via extended states. We have assumed that the electron spends $\tau$ seconds in the extended states before visiting a trap site. The average value of $\tau$ (about 2 fs) is very small when compared to the value of $\langle t_i \rangle$ (about 5 ps). [10, 24] Although small, considering the average time of transport via extended states plays a critical role when we incorporate the spatial occupation of localized states. According to the MT model the electron is immobile in the localized states and the transport mechanism is carried out solely by extended states (conduction band levels).[4] In this regard, two different times should be considered in the simulation of transport process (Figure 1): a time for electron residence in localized states (the first term of Eq. 7) and a time for transport in the extended states (the second term Eq. 7). Let us to see what will happen if we ignore $\tau$ in Eq. 7 during the simulation of electron transport. Figure 5 shows the results of such simulation. It is seen that by increasing Fermi level, the electron diffusion coefficient initially increases and then rapidly diminishes in the vicinity of mobility edge ($\sim 4\, kT$ below mobility edge). This result is expected since by rising Fermi level most of localized states will be occupied and the wanderer electron cannot transfer to them. As a consequence, there will be a threshold for the fraction of occupied sites for which the electron cannot find a percolation route. For our case (cubic lattice of sites and nearest neighbor transitions) this threshold of percolation is equal to the occupation of $\sim 70\ \%$ of total sites (compare Figure 4 and Figure 5).[25] As mentioned before, almost in all of the recent Monte Carlo simulations of MT model, the time of transport via extended states is ignored and such a threshold was not reported.[14, 16, 18, 19, 21] In the case of the conventional single-electron simulations, the effect of spatial occupation of traps is ignored and the electron always transport within an unoccupied lattice of trap sites (Figure 3.b and Ref. [14]). However, in the case of multi-electron simulation, such a behavior can be identified from the results of reference [14] (see figures 4 and 5 of reference [14]). Unfortunately, the multi-electron simulation is too time consuming at higher Fermi levels to allow us to simulate the electrons freezing in the absence of $\tau$. As a result, in the presence of spatial occupation of localized states it is necessary to consider the electron transport time in extended states.

All of the simulations were carried out in a 50×50×50 simple cubic lattice with a lattice constant of $a_{l0}$. In order to reduce the statistical fluctuations, the electron transport coefficient was obtained by averaging over 200 runs in all of the simulations.

## Numerical calculations

Combining equations 2, 5, and 6, the electron diffusion coefficient can be calculated as [14, 26]

$$D_J = \frac{1}{6}(1 - \alpha)exp\big(\varepsilon_f(1 - \alpha)\big)a_{l0}^2\nu_0 \qquad (9)$$

This is an approximate result for which the Fermi level is well below $E_l$ [14, 26]. It is shown that for moderately disordered media (0.2<α<0.4), the triple of single-particle simulation, many-particle simulation and the above analytical formula lead to almost same results. In the follows, we will derive an exact numerical formula for electron diffusion coefficient.

Let us to consider the motion of the electron when it moves through a bond (extended state) and reaches to the spatial position of a site or a localized state (see Figure 3.c). The corresponding traveled distance and time interval may be written as

$$r(t) = a_{l0} \qquad (10)$$

$$t = \tau + \langle t(E)\rangle \qquad (11)$$

where $\langle t(E)\rangle$ is the average of electron residence time in a trap site. Substituting equations 10 and 11 into the Eq. 6 we obtain

$$D_J = \frac{1}{6}\frac{a_{l0}^2}{\tau+\langle t(E)\rangle} \qquad (12)$$

rearrange above equation and define the normalized diffusion coefficient as

$$\widetilde{D}_J = \frac{D_J}{a_{l0}^2\tau^{-1}/6} = \frac{\tau}{\tau+\langle t(E)\rangle} \qquad (13)$$

Since the electron can only be trapped in an unoccupied site, we may calculate $\langle t(E)\rangle$ as

$$\langle t(E)\rangle = \frac{\int_{-\infty}^{0} g(E)t(E)\big(1-f_{FD}(E-E_f)\big)dE}{\int_{-\infty}^{0} g(E)dE} \qquad (14)$$

Substituting equations 1 and 5, the average time of electron residence in a trap site is obtained as

$$\langle t(E)\rangle = \nu_0^{-1}\eta \qquad (15)$$

where $\eta$ is a dimensionless parameter given by

$$\eta = e^{-2\varepsilon_f}\left(e^{\varepsilon_f} - \frac{\alpha}{1+\alpha}\,{}_2F_1(1.1+\alpha.2+\alpha.-e^{-\varepsilon_f})\right) \qquad (16)$$

Finally by substituting Eq. 15 in Eq. 13 and define another dimensionless parameter as $\xi = \tau\nu_0$ we get the normalized electron diffusion coefficient as

$$\widetilde{D}_J = \frac{\xi}{\xi+\eta} \qquad (17)$$

This equation separates two regimes for electron transport in a disordered medium. In the case of $\eta \gg \xi$ the transport is trap-limited and the normalized diffusion coefficient is given by $\widetilde{D}_J = \xi/\eta$. On the other hand, for the case of $\eta \ll \xi$ the electron transport is dominated by delocalized states and $\widetilde{D}_J = 1$. This result is similar to the famous multiple-trapping formula for electron effective mobility $\tilde{\mu} = \mu^*/\mu = \tau_c/(\tau_c + \tau_r)$ where $\tau_c$ and $\tau_r$ are the trapping lifetime the trap release time, respectively. [27]

It is worth noting that in the deriving Eq. 17 no assumption was used and we just rewrite the model mathematically. For compare this result with the Eq. 9 let us to multiply Eq. 17 by $\xi^{-1}$ and rewrite this equation as

$$D_J = \frac{1}{6}\left(\frac{1}{\xi+\eta}\right) a_{l0}^2 \upsilon_0 \qquad (18)$$

Figure 6 shows the normalized diffusion coefficient calculated according to the equations 9 and 18. It is seen that for typically lower values of $\alpha$ ($\alpha$<0.3) two equations are essentially identical at the entire interval of Fermi level. For higher values of $\alpha$, however, the results of two equations separate at higher Fermi levels (in the vicinity of mobility edge). As $\alpha$ increases, the diffusion coefficient obtained from Eq. 9 tends to lower values in the vicinity of mobility edge. This trend is in contrast of expected general behavior for electrons for which the transport facilitates as the value of $\alpha$ rises. As mentioned in reference [14], the Eq. 9 is derived under the assumption that the Fermi level is well below $E_l$. In comparison, Eq.18 which is derived without any approximation shows the expected behavior for electron transport at higher values of Fermi level.

The numerical Eq.18 reflects the exact nature of electron transport in a simple cubic lattice of localized states and in the subsequent we will use it for justification the MC models.

# III. RESULTS AND DISCUSSION

Figure 7 shows the temporal evaluation of electron diffusion coefficient for $\alpha$=0.2 and $\alpha$=0.6 simulated according to the model #1 and model #2. It is seen that in the region of dispersive transport the slopes of two models are identical. According to the results of model #2 for a strongly disordered medium ($\alpha$=0.2), the normal diffusion process begins at longer times. It can be seen from the right panel of Figure 7 that as $\alpha$ increases (less disorder medium) the difference between two models disappears. This observation is a direct consequence of the spatial occupation of trap sites. From the numerical point of view, there is no occupied site in the simulation lattice of model #1 and the electron always encounters with an unoccupied site. In this regard, the sampling rate from localized states is typically high and the thermal equilibrium with trap sites is obtained at shorter times. On the other hand, in the simulation lattice of model #2 a visited trap site can be either occupied or unoccupied leading to a lower sampling rate typically smaller than the model #1 and the thermal equilibrium is obtained at longer times in the model #2. Since at a fixed Fermi level the fraction of occupied states decreases when $\alpha$ rises (Figure 4), the equilibrium is obtained almost at the same time for two models in a moderately disordered medium.

Figure 8 represents the electron diffusion coefficient vs. normalized Fermi level simulated at different values of $\alpha$. It is seen that at the moderately disordered medium (α = 0.6), the results obtained from two models are identical and consistent with Eq. 18. In contrast, as the value of $\alpha$ decreases (being more disorder) the results of model #1 deviate considerably from numerical calculations. Similar deviation has been observed in the other groups simulations.[14, 17] This deviation is partially due to the fact that the model

#1 neglects the spatial occupation of localized states which is a function of Fermi level and disorder parameter. As indicated in Figure 4, the average distance between unoccupied states increases by reducing the disorder parameter. In addition, it is seen that model #1 overestimates the electron diffusion coefficient. The reason is that when the Fermi-Dirac distribution is replaced by a step function, the unoccupied deeper traps below the Fermi level are neglected and the additional occupied traps above the Fermi level are considered in the trapping/detrapping events leading to the average smaller transport time (see Figure 2). Since model #2 implies the raw DOLS and the spatial occupation of trap sites, these issues do not arise in this model and its results are quite consistent with the numerical calculations of Eq. 18. We repeat the simulations for other values of $\alpha$ and the results are summarized in the Figure 9. This figure represents the normalized difference in the diffusion coefficients obtained from model #1 and model #2. It is observed that the difference between two models exceeds 100% for $\alpha$<0.4.

Variation of electron diffusion coefficient with respect to the temperature is depicted in Figure 10. The Arrhenius character in both models can be verified from the linear behavior of transport coefficient vs. temperature.[28] It is seen that as the characteristic temperature of localized states increases, the results of model #1 become inconsistent with the numerical calculations. This implies that for strongly disordered media, the model #1 always predicts a smaller activation energy. In comparison, the results of model #2 remain consistent with the Eq. 18 at different values of $T_0$ and a broad interval of temperatures. We have fitted the data of Figure 10 employing the Arrhenius equation ($D_j = \kappa\, e^{-E_a/k_B T}$). The fitting results are summarized in Table 1 which confirm that the model #1 underestimate the activation energy. This is due to the employing the zero-temperature approximation in the model #1 which ignores the empty deeper localized states with higher activation energies (see Figure 2). It is worth noting that the results of two models converge at higher temperatures due to the exponential dependence of residence time on the temperature.

These results indicate that although the single-particle MC simulation based on the truncated DLOS is applicable for the moderately disordered materials like nanocrystalline anatase $TiO_2$ [29-31], it cannot be used for strongly disordered systems such as ZnO where the $\alpha$ values are very low and range between 0.05 and 0.15.[32-34] The model #2, on the other hand, can be used for any disordered medium independent of the value of α.

# IV. CONCLUSION

It was shown that in the presence of strong disorder, the conventional single-particle Monte Carlo simulation of multiple-trapping transport which is based on the truncated density of localized states deviates considerably from numerical calculations. The origins of the deviations were discussed on the bases of zero-temperature approximation and the spatial occupation of localized states. To eliminate these issues, we proposed a modified single-particle Monte Carlo simulation which uses the raw density of localized states and a modified electron transport time. The proposed model can reproduce the exact numerical results at a broad interval of disorder parameters, Fermi levels, and temperatures. This model can also be used for hopping transport where the distance between occupied and unoccupied states directly affects the electron transition rates.

# Acknowledgment

The authors wish to thank the Iran National Science Foundation (INSF) for partial financial support. Partial financial support from the Centre of Excellence on the Structure of Matter of the University of Tehran is also acknowledged.

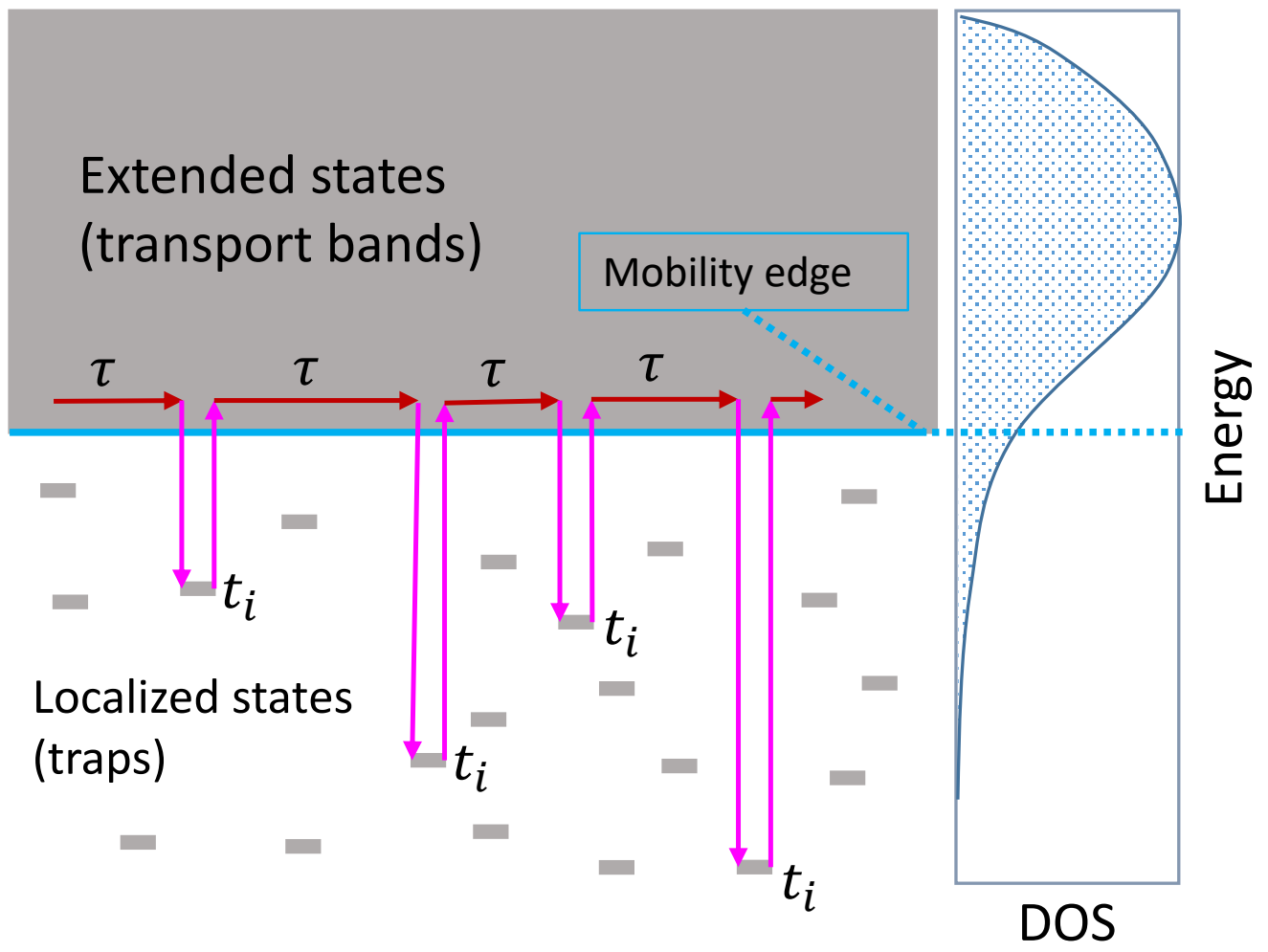

Figure 1 Schematic of multiple-trapping transport. The transport is carried out via extended states while the electron is immobile in the localized states. $\boldsymbol{\tau}$ is the average time that the electron spends in the extended states between two successive trapping events. The electron residence time in a localized state ($t_i$) is given by Eq.5.

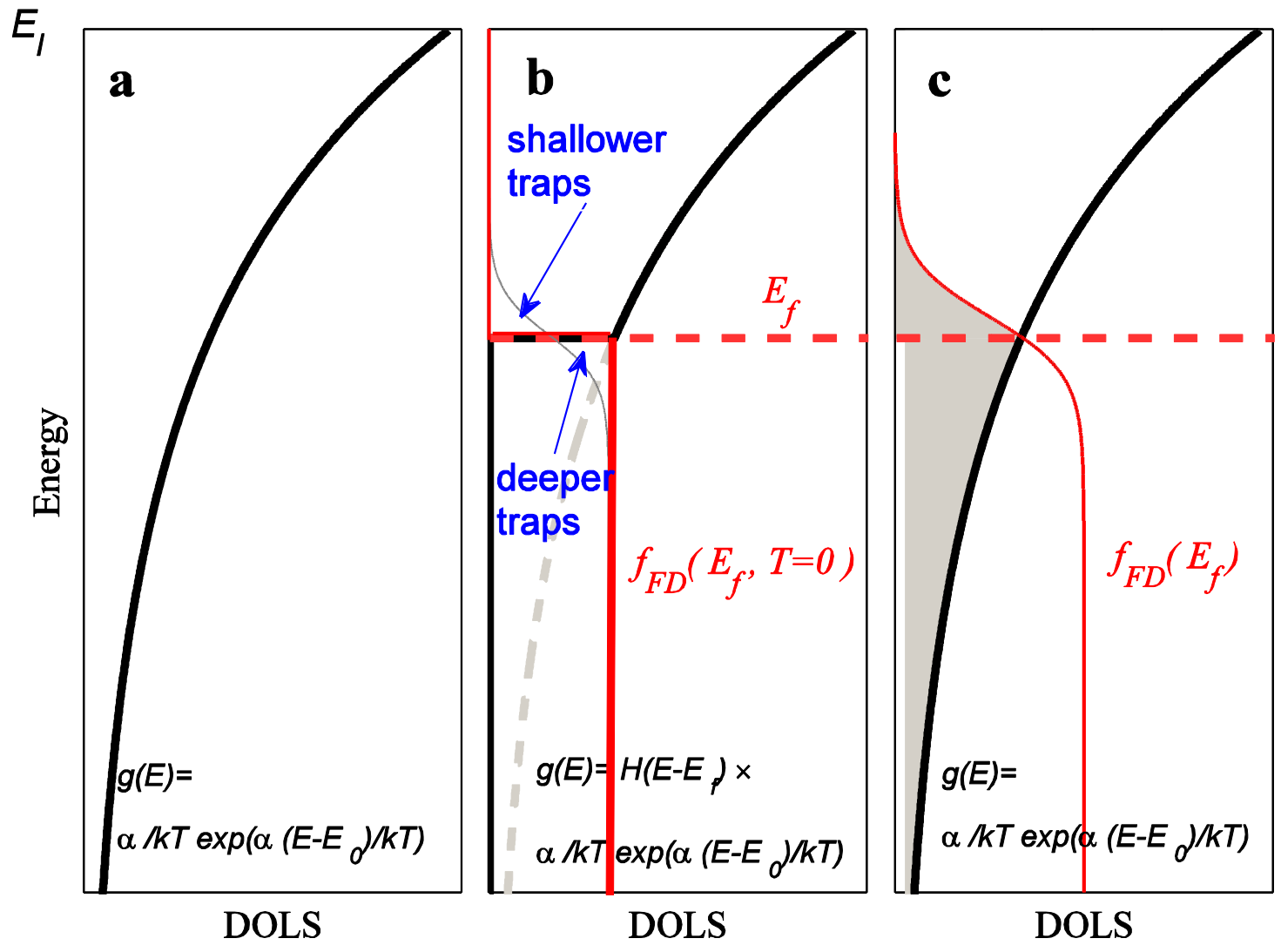

Figure 2. (a) Exponential density of localized states. (b) Truncated density of localized states used in the conventional single-particle MC model. (c) Schematic of modified single-particle MC model where the gray region indicates the occupied states.

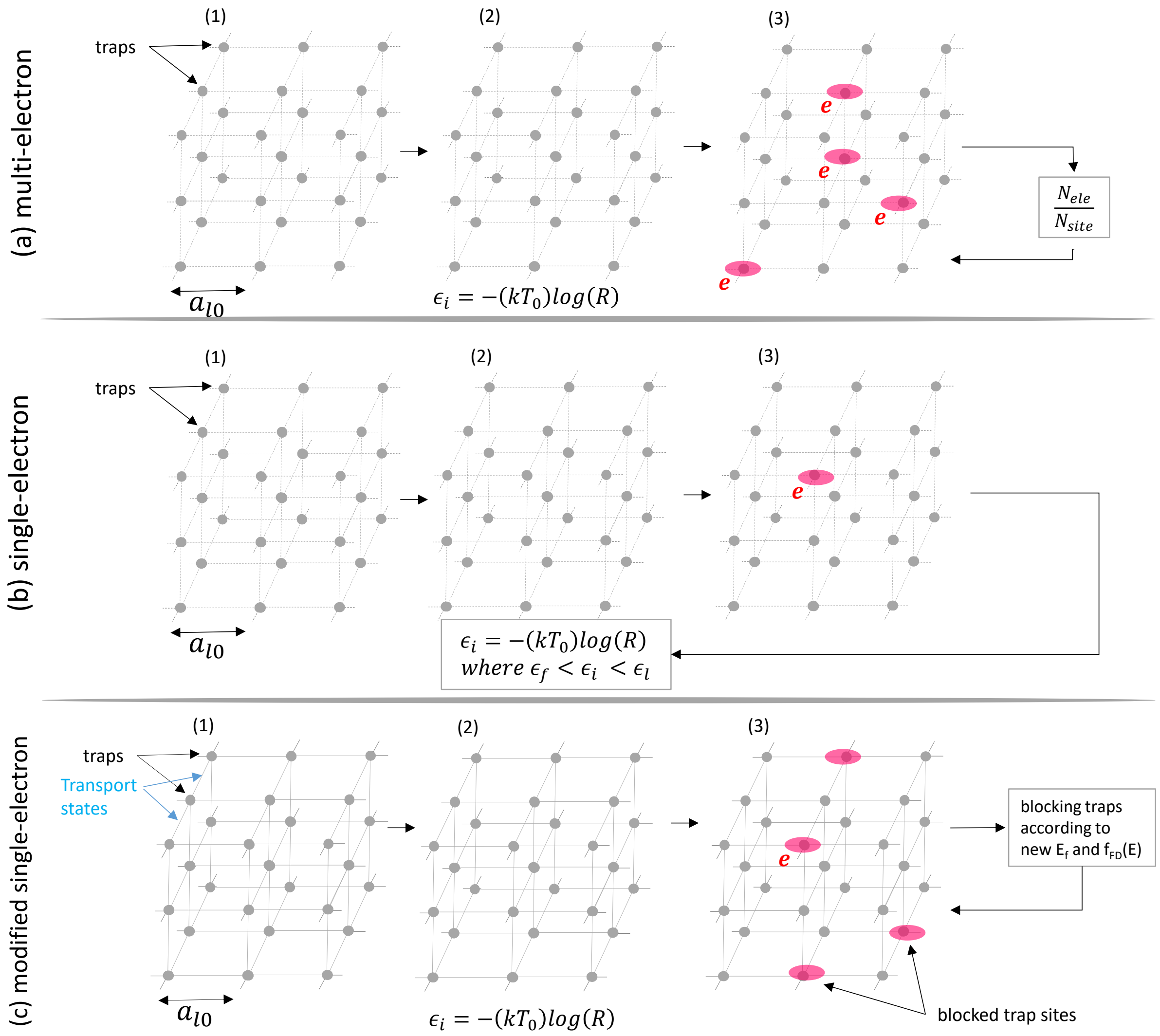

Figure 3. Typical steps for the simulation of charge transport in a disordered medium: (*step I*) spatial distribution of localized states, (*step II*) distribution the energies of localized states, and (*step III*) the random walk process. (a) Multi-electron: tfhe Fermi level is tuned via changing the number of electrons in *step III*. (b) Conventional single-electron: according to the Fermi level the energies of trap sites are redistributed (*step II*). (c) Modified single-electro: each site has an occupation index that is updated for new Fermi level according to the Fermi-Dirac distribution. The blocked sites indicate the occupied localized states and are not available for trapping the test electron. Moreover, in this model the transport via delocalized states is also considered in the transport process (bonds in *step I* indicated by transport states).

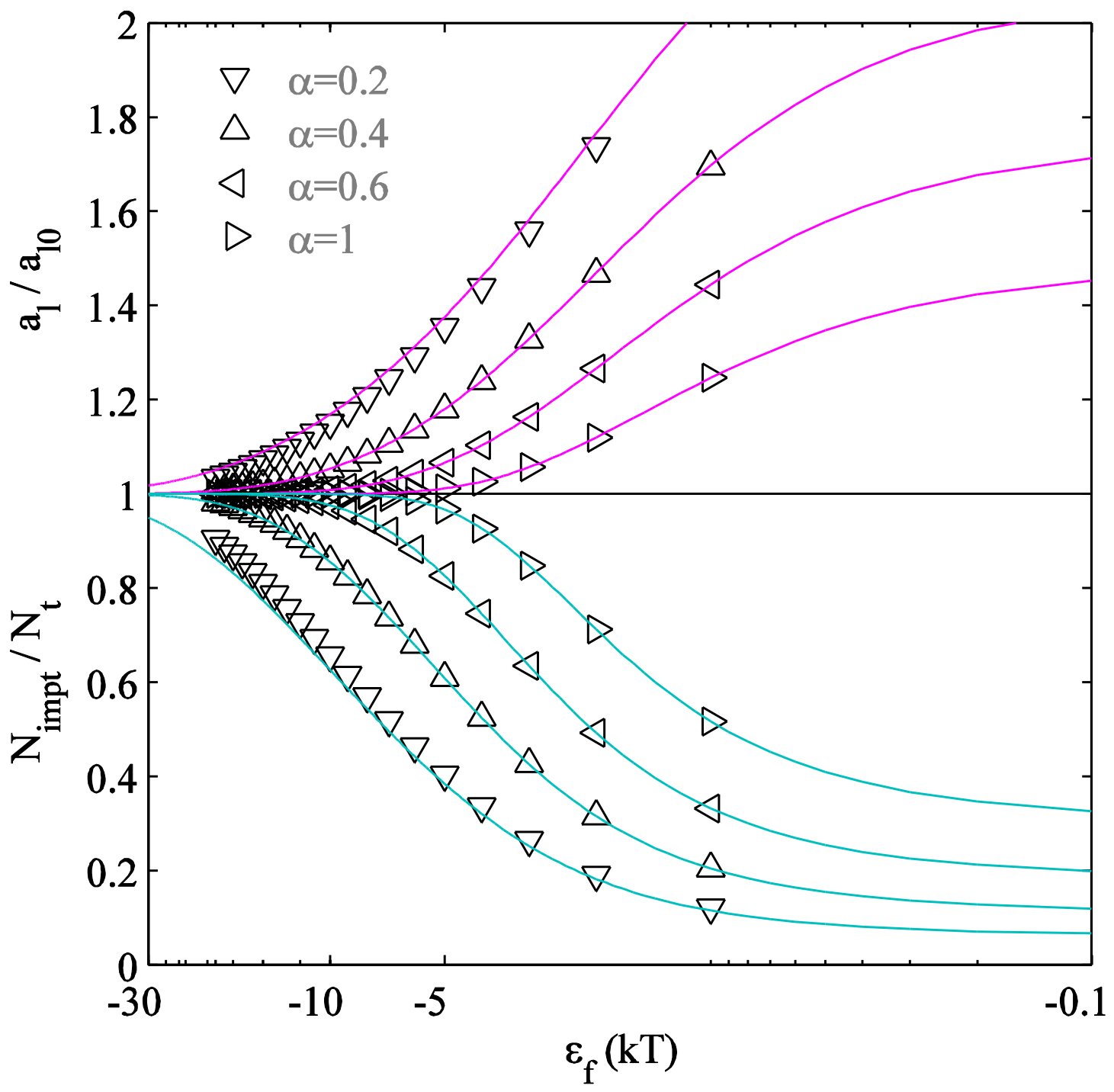

Figure 4. The average distance between unoccupied sites and the fraction of unoccupied states as a function of Fermi level. Marks and lines represent the results of Monte-Carlo simulation and Eq. 4 respectively. $a_{l0}$ is the lattice constant of simple cubic lattice.

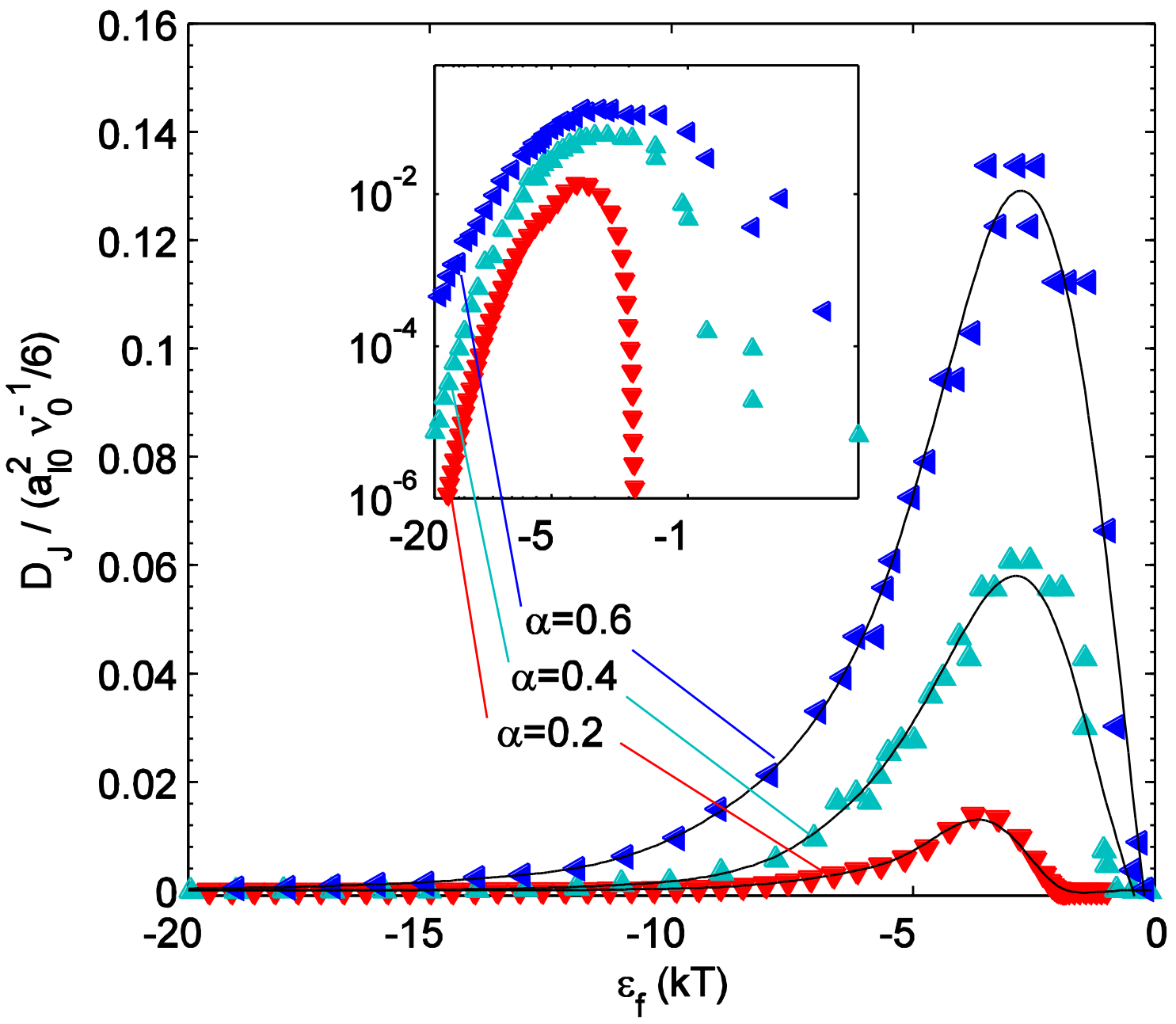

Figure 5. Electron diffusion coefficient vs. normalized Fermi level simulated in the framework of model #2 and neglecting the transport time via extended states (i.e. $\tau$ in Eq. 7). The inset panel shows the same results in logarithmic scale.

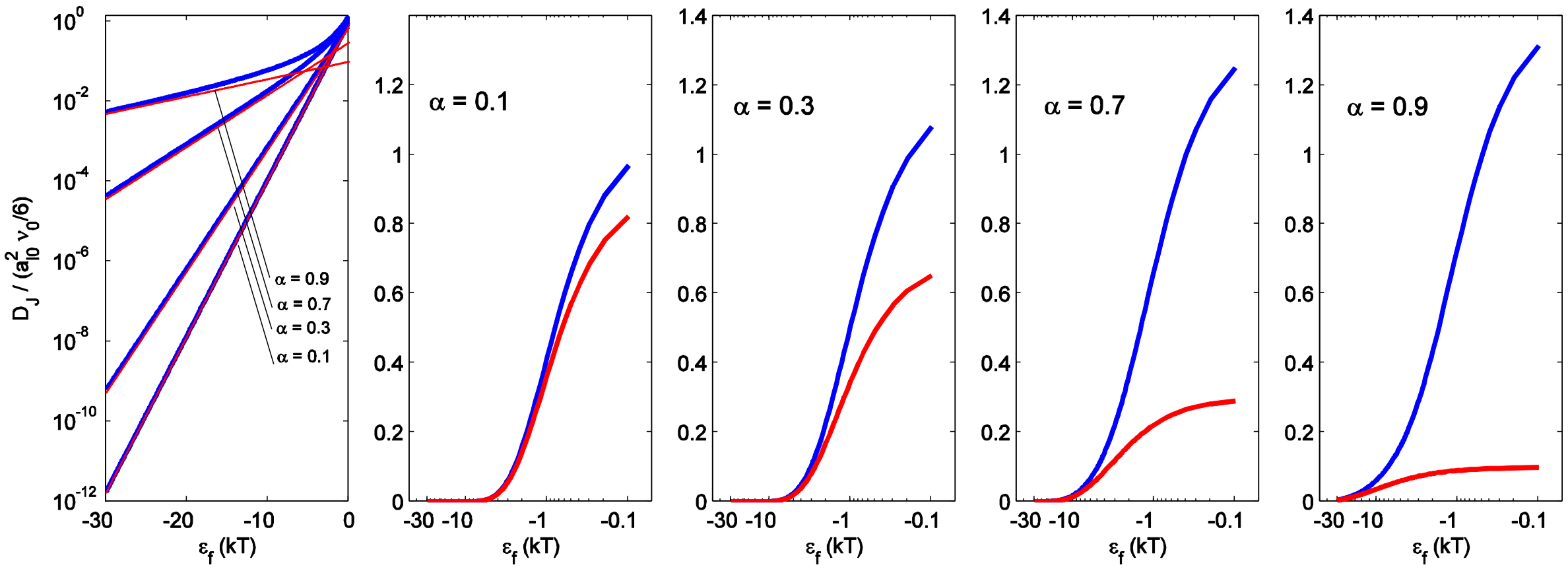

Figure 6. Normalized electron diffusion coefficient calculated according to Eq.9 (red lines) and Eq.18 (blue lines).

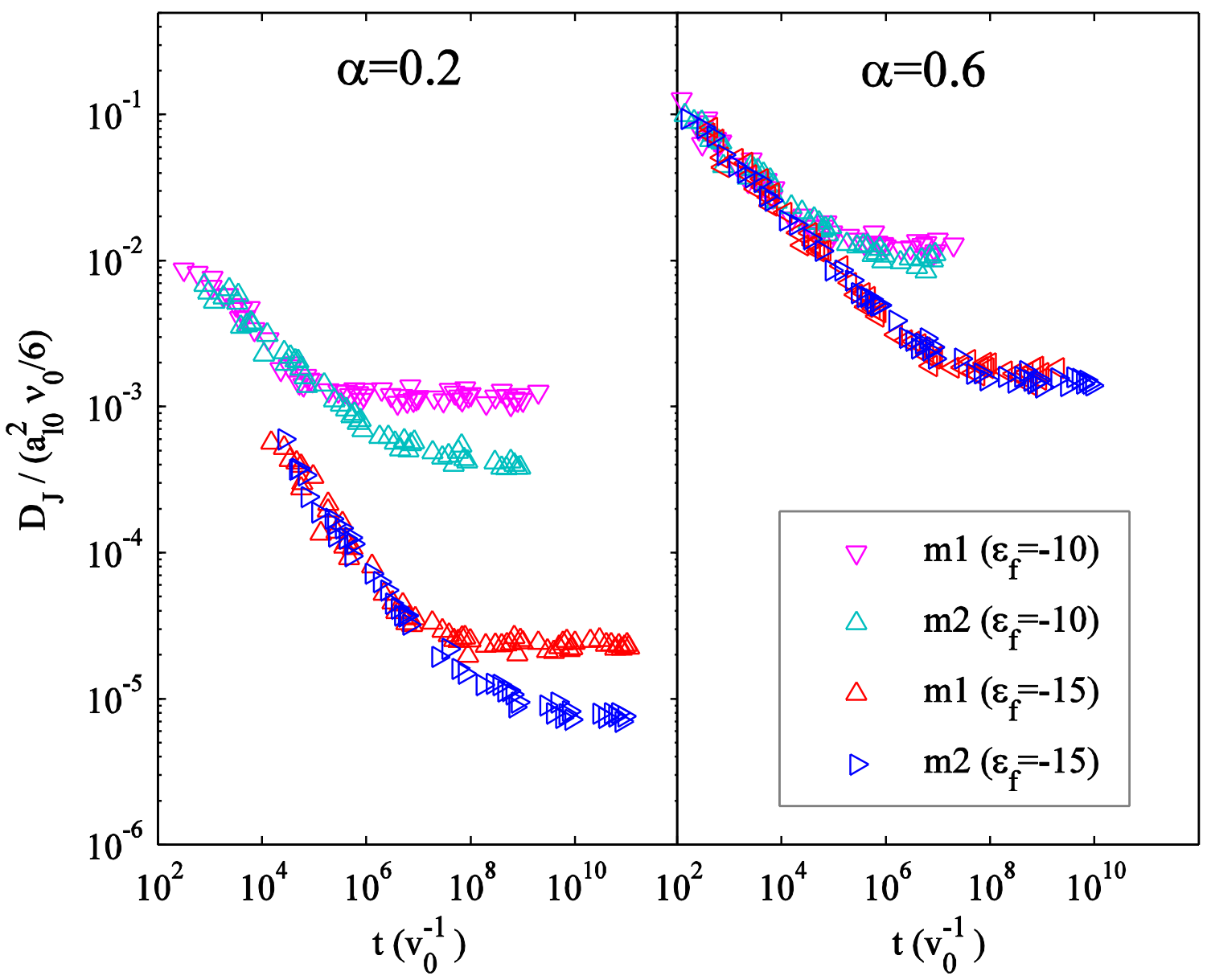

Figure 7. Temporal evaluation of electron diffusion coefficient obtained from simulations based on model #1 (m1) and model #2 (m2) at $T = 300\ K$ and $\xi = \tau v_0 = 10^{-4}$ . The non-dispersive (normal) transport is defined as a region for which $dD/dt = 0$. The error bars fall within the marker size.

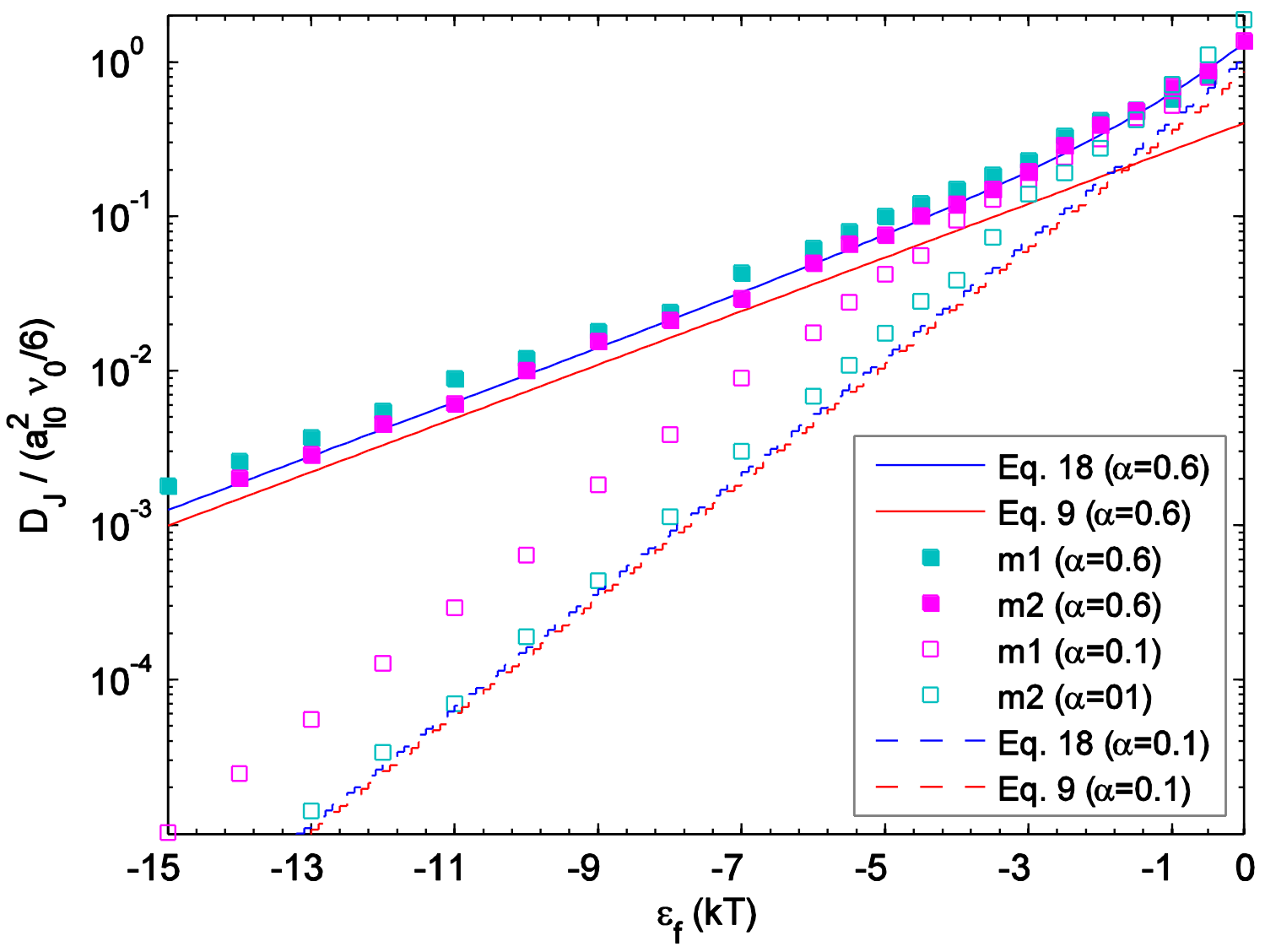

Figure 8. Electron diffusion coefficient vs. normalized Fermi level. All of the error bars of Monte Carlo simulations fall within the marker size.

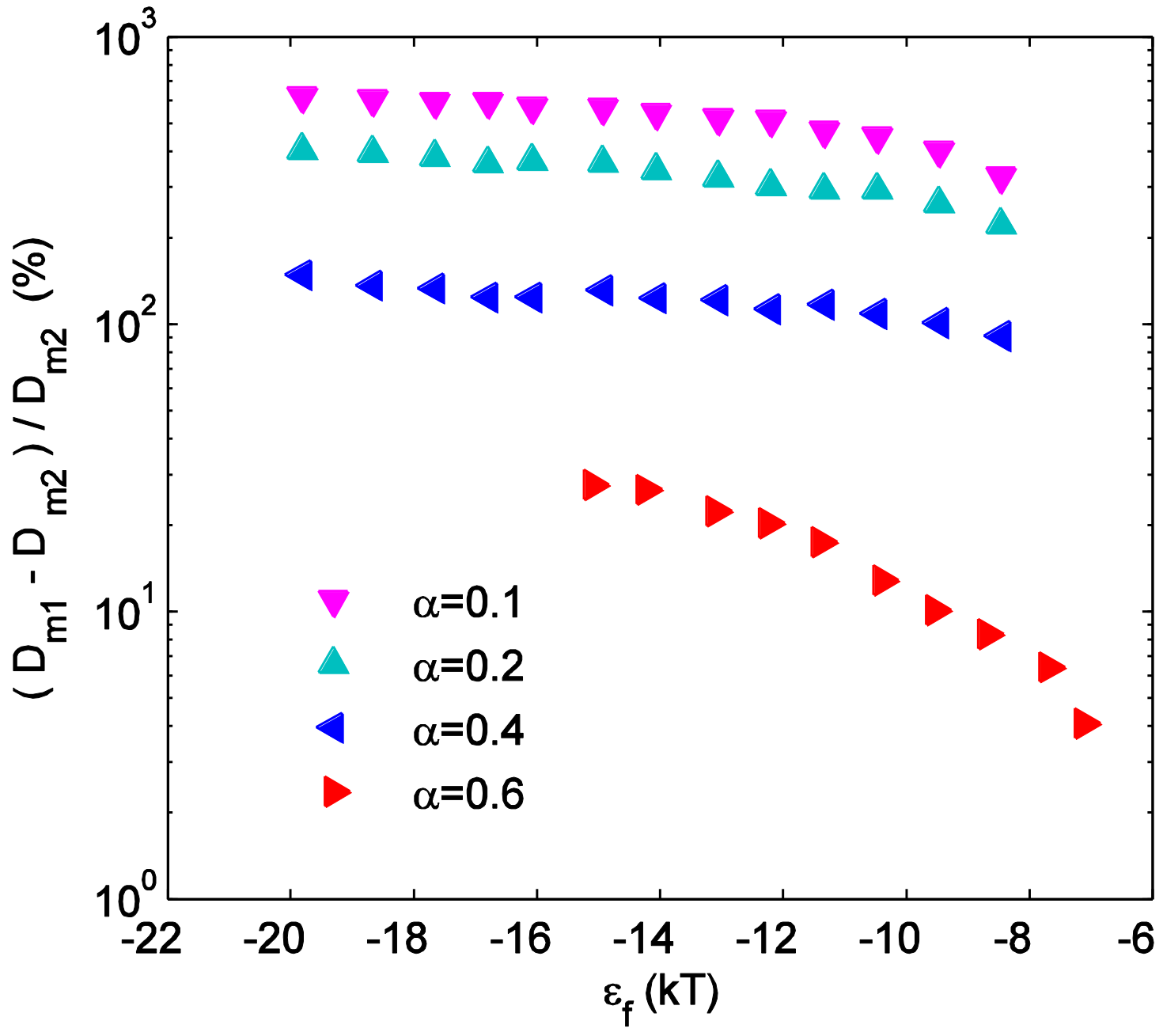

Figure 9. The normalized difference between diffusion coefficients obtained from model #1 and model #2.

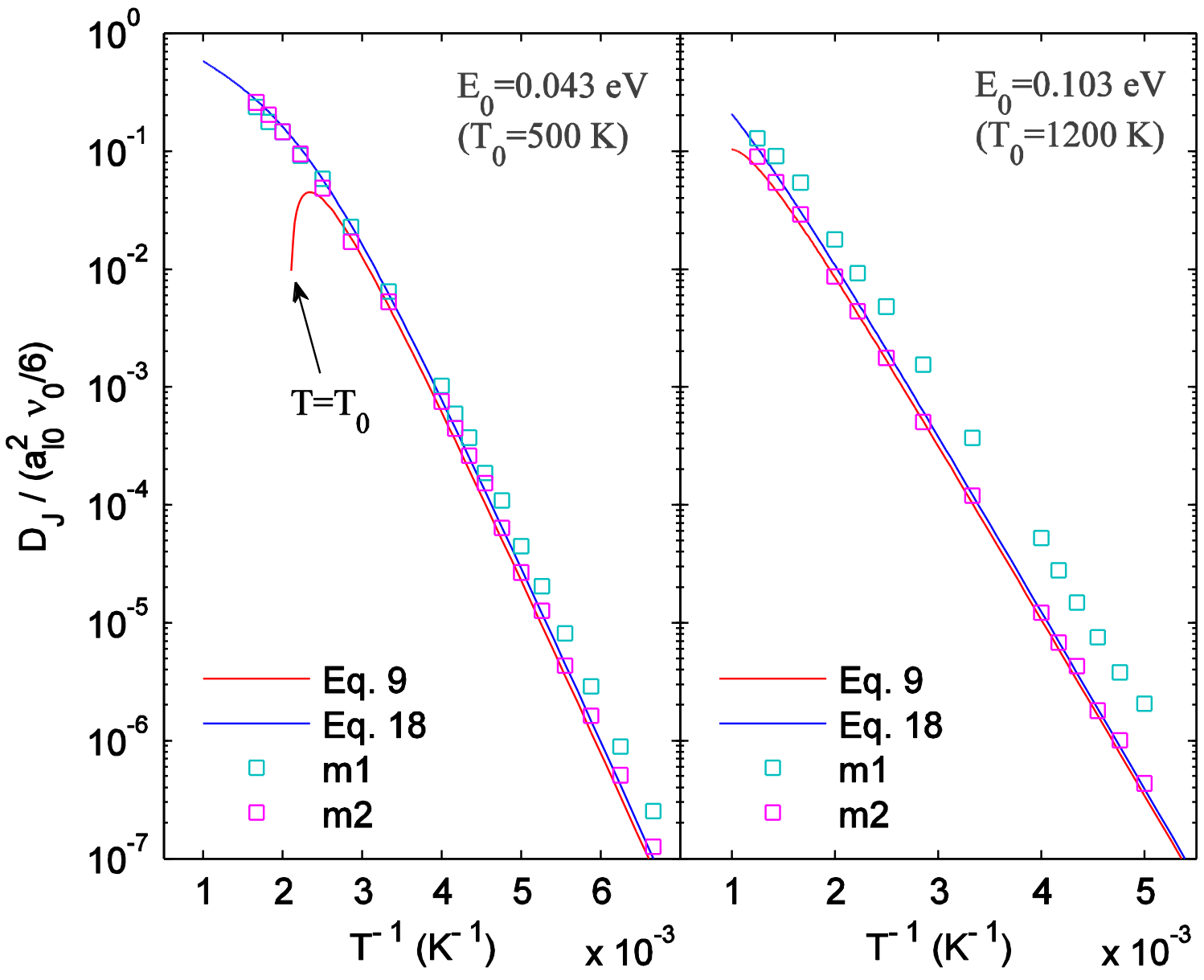

Figure 10. Electron diffusion coefficient vs. temperature at constant Fermi level -0.3 eV.

Table 1 fitting parameters for the Arrhenius plots (Figure 10)

| $T_0$ (K) | | m1 | m2 | Eq. 9 | Eq. 18 |
|---|---|---|---|---|---|
| 500 | $\kappa\ (a_{l0}^2 \upsilon_0/6)$ | 46.70 | 158.64 | 106.45 | 134.41 |
| | $E_a(eV)$ | 0.240 | 0.268 | 0.268 | 0.268 |
| 1200 | $\kappa\ (a_{l0}^2 \upsilon_0/6)$ | 6.33 | 4.93 | 7.00 | 3.82 |
| | $E_a(eV)$ | 0.257 | 0.278 | 0.287 | 0.278 |